\documentclass[preprint,11pt]{elsarticle}

\usepackage{graphics}
\usepackage{graphicx}
\usepackage{verbatim}
\usepackage{acronym}
\usepackage{color}
\usepackage{comment}
\usepackage{subcaption}
\usepackage[T1]{fontenc}
\usepackage[english]{babel}
\usepackage{amsmath,amsfonts,amsthm}
\usepackage{amssymb}
\usepackage[linesnumbered,ruled]{algorithm2e} 
\usepackage{csquotes}
\usepackage{url}
\usepackage{paralist}
\usepackage{verbatimbox}
\usepackage{verbdef}
\usepackage{times}

\newacro{MVDA}{Multivariate Data Analysis}
\newacro{EPLS}{Exploratory Projection to Latent Structures}
\newacro{RMSE}{Root Mean Squared Error}
\newacro{SVD}{Singular Value Decomposition}
\newacro{PCA}{Principle Component Analysis}
\newacro{ANOVA}{ANalysis Of VAriance}
\newacro{mRNA}{messenger Ribonucleic Acid}
\newacro{string-db}{STRING Database~\footnote{https://string-db.org/}}
\newacro{MNIST}{Modified National Institute of Standards and Technology}
\newacro{GSE2508}{\emph{Expression profiling in adipocytes of obese humans}}
\newacro{DBSCAN}{Density-Based Algorithm for Discovering Clusters}
\newacro{UMAP}{Uniform Manifold Approximation and Projection}
\newacro{CCA}{Connectivity Clustering Algorithms}
\newacro{LCA}{Link Clustering Algorithms}
\newacro{AHC}{Agglomerative Hierarchical Clustering}
\newacro{GRS}{Golden Ratio Search}
\newacro{GSS}{Golden Section Search}
\newacro{RNA}{Ribonucleic Acid}
\usepackage{listings}
\usepackage{hyperref}
\makeatletter
\def\ps@pprintTitle{%
  \let\@oddhead\@empty
  \let\@evenhead\@empty
  \def\@oddfoot{\reset@font\hfil\thepage\hfil}
  \let\@evenfoot\@oddfoot
}
\makeatother
\begin{document}

\begin{frontmatter}

\title{ Clustering Optimisation Method for Highly Connected Biological Data }

\author[kth,sfl]{Richard Tj\"{o}rnhammar}
\ead{richardt@kth.se}

\address[kth]{KTH Royal Institute of Technology, SE-100 44 Stockholm, Sweden}
\address[sfl]{SciLifeLab, Tomtebodavägen 23, SE-171 65 Solna, Sweden}
\begin{abstract}

Currently, data-driven discovery in biological sciences resides in finding segmentation strategies in multivariate data that produce sensible descriptions
of the data. Clustering is but one of several approaches and sometimes falls short because of difficulties in assessing reasonable cutoffs, the number of 
clusters that need to be formed or that an approach fails to preserve topological properties of the original system in its clustered form. In this work, 
we show how a simple metric for connectivity clustering evaluation leads to an optimised segmentation of biological data.

The novelty of the work resides in the creation of a simple optimisation method for clustering crowded data. The resulting clustering 
approach only relies on metrics derived from the inherent properties of the clustering. The new method facilitates knowledge for optimised
clustering, which is easy to implement.

We discuss how the clustering optimisation strategy corresponds to the viable information content yielded by the final segmentation. We further
elaborate on how the clustering results, in the optimal solution, corresponds to prior knowledge of three different data sets.

\end{abstract}


\end{frontmatter}

\section{Introduction}

One key feature of biological data is the excessive amount of viable information. Real-world systems describing a simple cellular or 
biochemical process are often large, containing many active reagents in a crowded 
environment~\cite{kanehisa_kegg_2021,ashburner_gene_2000,gillespie_reactome_2022,szklarczyk_string_2021}. Depending on the nature of the interactions 
belonging to the constituents of such a system various dimensionality reduction techniques can be employed to coarse grain the system and reduce 
the complexity of the studied problem~\cite{pedregosa_scikit-learn_2011,mcinnes_umap_2018}. Determining an optimal number of clusters for more in-depth 
analysis or visualisation is an open problem with many 
different solutions~\cite{liu_multik_2021,kiselev_sc3_2017,satija_spatial_2015,ronen_deepdpm_2022}. Here, we present a simplified optimisation 
approach and algorithms to achieve this task.

We employ clustering with a naming scheme where clustering segmentation produced via 'connection clustering', see methods, is referred to as 
\ac{CCA} and 'linkage' based clustering, commonly employed in \ac{AHC}~\cite{ward_hierarchical_1963}, is referred to as \ac{LCA}~\cite{sibson_slink_1973}. 
\ac{CCA} constitutes a point evaluation of the system 
distance matrix for a single distance cutoff ($D_{ij},\epsilon \in \mathbb{R}^+$ and $D_{ij}=D_{ji}$) while \ac{LCA} evaluates the linkage matrix 
describing the entire hierarchy. Here $D_{ij}$ describes all the pairwise distances between the parts of the entire system. \ac{CCA} will let you determine 
if distance matrix indices are connected at some distance or not. The connection-based methods establish the number 
of clusters in the binary Neighbour matrix without constructing an intermediate linkage matrix. 

The Neighbour matrix is defined here as the pairwise distance between the parts $i$ and $j$ of the system ($D_{ij}$) with an applied cutoff 
($N_{ij} = D_{ij}\le\epsilon$) and is related to the adjacency matrix from graph theory by adding an identity matrix to the adjacency matrix 
($A_{ij} = N_{ij} - I_{ij}$). The three boolean matrices that describe a system at some distance cutoff ($\epsilon$) are: 
the Identity matrix ($I_{ij}=D_{ij}\equiv0$), 
the Adjacency matrix ($A_{ij}=D_{ij}\le\epsilon - I_{ij}$) and 
the Community matrix ($O_{ij}=D_{ij}>\epsilon$). We note that summing the three matrices will return $1$ for any $i,j$ pair. \ac{CCA} determines
the number of clusters by traversing $N_{ij}$ and evaluates if there is any true overlap for a specific distance cutoff. Publically 
available \ac{CCA} methods include the connectivity Algorithm~\ref{alg:connectivity} in the methods section as well as 
the \ac{DBSCAN}~\cite{10.5555/3001460.3001507,pedregosa_scikit-learn_2011} without point rejections.

Linkage algorithms determine the number of clusters for all unique distances by forming the linkage matrix reducing and ignoring some connections
to already linked constituents of the system in accord with a chosen heuristic. A Linkage algorithm is not an unambiguous treatment of a 
system~\cite{fernandez_solving_2008} where all the true connections in it are important, such as in a molecular water bulk
system, when you want all your quantum-mechanical water molecules to be treated at the same level of theory based on their connectivity at a specific 
distance. If you are doing statistics on a complete hierarchy then this distinction is not important. You can construct hierarchies from both algorithm 
types but a connection algorithm, without point rejection criteria, will always produce a unique and well-determined structure while the link algorithms 
will be unique but structurally dependent on how ties are resolved and which heuristic is employed for construction. The connection hierarchy is exact 
and unique, but slow to 
construct, while the link hierarchies are heuristic dependent, but fast to construct. The Linkage algorithms are more efficient at creating a hierarchy 
based on a distance matrix representation of the data but can be thought of as throwing away information at every linking step. The full link 
algorithm determines the new cluster distance to the rest of the unclustered points in a self-consistent fashion by employing two different heuristics. 
Minimal distances for assigning a cluster link are determined by finding the minimum
non-diagonal element in the distance matrix and the new link group distance to all remaining points are determined using the second heuristic. 
Using simple linkage, or $\min$ value distance assignment, ensures that the same heuristic is employed both when finding links as well as 
when assigning the link group to system
distances. We will see that it will also produce an equivalent clustering as compared to the one deduced by a connection algorithm for a 
specific distance $\epsilon$. Except for some of the cases when there are distance ties in the link evaluation. This is a computational quirk that does 
not affect 'connection' based hierarchy construction.

For a specific $\epsilon$ we have $C$ cluster segments ($\dim(C)=c$ and $c \in \mathbb{Z}^+$) with $K_i$ parts in each for a system comprised of $N$ parts. 
Furthermore, for a given minuscule $\epsilon$, the clustering segmentation is exactly all the parts of the system ($\dim(K_i)=1$, $c=N$).
In the same fashion for a huge $\epsilon$ the entire system resides in a single cluster ($\dim(K_0)=N$, $c=1$). It is clear that the number of clusters
describing the system is a monotonically decreasing function of $\epsilon$ while the average size of the clusters is a monotonically increasing function 
of $\epsilon$. We define the two functions as 
\begin{equation}
M(\epsilon) = c(\epsilon)
\label{eq:M}
\end{equation}
for the decreasing function and
\begin{equation}
\overline{S(\epsilon)} = <\dim(K_i)>_C =  \frac{1}{c(\epsilon)} \sum_{i=0}^{c(\epsilon)} \dim(K_i).
\label{eq:S}
\end{equation}
for the increasing function.

To deduce an informative clustering cutoff $\epsilon$ we form the function $G$ defined as
\begin{equation}
G(\epsilon) = \frac{ \overline{S(\epsilon)}M(\epsilon) }{ \overline{S(\epsilon)} + M(\epsilon) } - \frac{N}{N+1}
\label{eq:G}
\end{equation}
We note that both functions ($\overline{S} \in [1.0 \cdots N]$, $M \in [N \cdots 1.0]$) contain complementary information and that $G$ obtains 
the maximal value inside the interval.
Since the $G$ function is unimodal we can conduct a \ac{GRS}~\cite{kiefer_sequential_1953} to find the optimal $\epsilon$.
Changing the type of $S$ function into any quantile value function, a $\min$ or $\max$ function should 
not change the unimodality of the geometric $G$ function. If we chose 
to employ a $\min$ ($S^-$) value function then the optimum should be pushed to a larger $\epsilon$ value while the $\max$ valued ($S^+$) should in general 
yield a smaller $\epsilon$ solution as compared to the mean. For highly structured data, with an extremely peaked distribution of pairwise distances, 
these assumptions are not true. One such example is an ideal $2D$ graphene mesh. The ideal graphene bond distances are the same for every connection 
in the hexagonal mesh. The $S$ heuristic function will be a Heaviside function jumping from $1$ to $N$ at a single distance while the $M$ will have 
the opposite behaviour. This is symmetric for the entire system and will thereby yield a $G$ function response without structure and not exhibit 
any modality. For systems with a small persisting cluster of only a single constituent until the very last step will result in a unimodal $G,min$, 
but the mode will be a global minimum.

To calculate a compositional specificity metric of a cluster to non-binary multiclass labelled targets belonging to analytes in $K_i$ with label counts $x$, 
we employ a metric defined by :
\begin{equation}
\gamma = \frac{x^+}{\sum_i{x_i}} \cdot (1-e^{-(\dim{(x)}-1)} )\quad ,\quad \dim{(x)} \geq 1
\label{eq:specificity}
\end{equation}
Another such metric is the $\tau$ specificity~\cite{kryuchkova-mostacci_benchmark_2016}, but we refrain from using it since it is not well defined for 
clustering solutions where there exist clusters with only one part. Here $x^+$ corresponds to the maximum value of the label counts in the vector $x$.

It is clear that a linkage method is more efficient for constructing complete agglomerative hierarchies while a single `connectivity` 
the calculation might be more efficient if you only want the clusters at a predetermined distance. Searching for an optimal distance cutoff for the cluster
representation will also be heuristic dependent. However using this approach, with $S$ and $M$ functions as \ac{CCA} heuristics for the $G$ metric, finding 
an optimal $\epsilon$ becomes a unimodal optimisation problem.

For completeness, we will include the description of an exhaustive connectivity method (Algorithm~\ref{alg:connectivity}) in the methods section as well
as a description of a \ac{GRS}~(Algorithm~\ref{alg:grs}).

\section{Method} \label{sec:method}

For any data set we first construct the distance matrix using a distance measure and data axes, such as a euclidian distance or a spearman 
correlation distance~\cite{solo_pearson_2019} between all parts of interest in the data set. For molecular water, this is usually the euclidian 
distance between the atomic positions. For a microarray dataset, it can be a correlation ($\rho$) distance 
($d=\sqrt{1-\rho}$) between the transcripts sample positions. In this work, we have chosen to employ euclidian pairwise distance metrics for all the 
studied data sets.

Given that there exists a segmentation of the system decomposed of $c$ clusters so that the entire system is encoded 
into $K$ cluster parts in accord with Algorithm~\ref{alg:connectivity}:~\ac{CCA} \footnote{ The \ac{CCA}, Connectivity algorithm 
have been implemented by the author in the publically available Python package 'impetuous-gfa'} then 
Algorithm~\ref{alg:grs}:~\ac{GRS}\footnote{ One such search 
function has been implemented by the author in the publically available Python 'impetuous-gfa' package in the 'optimisation' module.} ensures that we 
always find the optimum $\epsilon$ for a unimodal function. 
We use the heuristics of the \ac{CCA} given by Equations~\ref{eq:M}~and~\ref{eq:S} which are transformed into a unimodal function by Eq.~\ref{eq:G}.

We consistently employ an algorithm annotation scheme where subscripts denote index positions in the tensor of interest and where $\dim{(A)}$ or 
$\dim{(A_i)}$ denotes the number of elements in $A$ along its first axis.

\begin{algorithm}
    \caption{\ac{CCA}, Connectivity }
    \label{alg:connectivity}
    \LinesNumbered
    \SetKwInOut{Input}{input}
    \SetKwInOut{Output}{output}
    \Input{ A Distance matrix $D_{ij}$, a float $\epsilon$ }
    \Output{ list of cluster sizes $Q$ , cluster towards part index list $R$ }
    \SetKwBlock{Begin}{begin}{} 
    \SetAlgoLined
    \Begin{
        $L \leftarrow \dim{(D_{0j})} $ \\
        $R$, $w$, $P$, $Q$, $I$ are empty lists \\
        $C_0 \leftarrow 0$ \\
        \For {$i\in[0,L)$} {
            $w_{i} \leftarrow i+1 $ \\
	    $R_{2\cdot i  } \leftarrow 0$ \\
            $R_{2\cdot i+1} \leftarrow 0$ \\
            $I_i \leftarrow i $ \\
        }
        \While { $\dim{(I)}>0$ } {
            $i \leftarrow I_{\dim{(I)}-1}$ \\
            $I \leftarrow I_{j \in[0,\dim{(I)}-1) }$ \\
            $P \leftarrow $ empty list \\
            \If{ $w_{i}>0$ } {
                $C_0 \leftarrow C_0-1 $\\
                \For{ $j\in[0,L)$ } {
                    \If{$D_{ij}\le\epsilon$}{
                        $P_{\dim{(P)}}\leftarrow j$ \\
                    }
                }
                \While{$\dim{(P)}$>0}{
                    $k \leftarrow P_{\dim{(P)}-1}$\\
                    $P \leftarrow P_{j \in[0,\dim{(P)}-1) }$ \\
                    $w_{k}\leftarrow C_0$ \\
                    \For{$j\in[0,L)$}{
                        \If{$D_{ij}\le\epsilon$}{
                            \For{$q\in[0,L)$}{
                                \If{$w_{q}=j+1$}{
                                    $P_{\dim{(P)}}\leftarrow q$ \\
                                }
                            }
                        }
                    }
                }
            }
        }
    }				%
\end{algorithm} 		%
\begin{algorithm}		%
    \LinesNumbered		%
    \setcounter{AlgoLine}{37}   %
    \SetKwBlock{Begin}{}{end}	%
    \Begin{			%
        \For{$i\in[0,-1\cdot C_0)$} {
            $Q_i\leftarrow 0$\\
        }
        \For{$q\in[0,L)$}{
            $R_{2\cdot q+1}\leftarrow q$ \\
            $R_{2\cdot q  }\leftarrow w_{q}-C_0$ \\
            $Q_{R_{2\cdot q}} \leftarrow Q_{R_{2\cdot q}}+1$
        }
    }
\end{algorithm}

\begin{algorithm}
    \caption{\ac{GRS}}
    \label{alg:grs}
    \SetKwInOut{Input}{input}
    \SetKwInOut{Output}{output}
    \Input{ A Distance matrix $D_{ij}$, a float $\epsilon$ , a float $tolerance$  }
    \Output{ float $\epsilon_{optimal}$ }
    \SetKwBlock{Beginn}{beginn}{ende}
    \Begin{
        $a \leftarrow \min{(D_{ij})}$ \\
        $b \leftarrow \max{(D_{ij})}$ \\
        $\psi \leftarrow \frac{ \sqrt{5}-1 }{2}$ \\
        $c \leftarrow b - \psi \cdot (b-a)$ \\
        $d \leftarrow a + \psi \cdot (b-a)$ \\
        \While { $ d-c > tolerance $ } {
            $fc \leftarrow ($ $G $( heuristics(\ac{CCA}($D_{ij},c$)) ) )$^2$ \\
            $fd \leftarrow ($ $G $( heuristics(\ac{CCA}($D_{ij},d$)) ) )$^2$ \\
            \eIf { fc>=fd } {
                 $b \leftarrow d$ \\
                 $d \leftarrow c$ \\
                 $c \leftarrow b - \psi\cdot (b-a)$ \\
            }{
                 $a \leftarrow c$ \\
                 $c \leftarrow d$ \\
                 $d \leftarrow a + \psi\cdot (b-a)$ \\
            }
            $\epsilon_{optimal} \leftarrow \frac{c+d}{2}$ \\
        }
    }
\end{algorithm}

We have chosen three data sets. The first is a small molecular water system in the liquid state~\cite{allen_computer_1994} containing $32$~H$_2$O 
and a single Hydronium ion. The system is a single time frame of a CPMD~\cite{car_unified_1985} water 
simulation\footnote{http://www.theochem.ruhr-uni-bochum.de/~legacy.akohlmey/files/32spce-h3op-1ns.xyz}. The second set contain microarray 
transcript readings of adipocytes with accession information: \ac{GSE2508}, 
that employed the $GPL8300$ platform, describing $20$ obese and lean men and women~\cite{lee_microarray_2005}. The third data is the first 
$35000$~\ac{MNIST}~digit images~\cite{lecun_backpropagation_1989}\footnote{http://yann.lecun.com/exdb/mnist/}. 

For all three data sets the full \ac{AHC} solutions, employing single linkages, were also computed for comparison. The water coordinates 
were used as is while the two larger data sets were processed by transforming to standardised values, by removing the mean and dividing with 
the standard deviation across samples or pixels. The microarray and digits were further transformed using 
\ac{UMAP}~\cite{mcinnes_umap_2018} prior to clustering.

The microarray transcripts were also analysed using a two-way \ac{ANOVA}~\cite{seabold2010statsmodels,gelman_analysis_2005} modulation for the body 
type class and biological sex. The significance for body type p-values was employed while sex was considered as a blocking variable. The generated 
p-values were adjusted using q-value rank correction~\cite{storey_statistical_2003} and used as input to calculate cluster significances. Cluster 
significances were determined by using a Fisher exact test~\cite{fisher_interpretation_1922} where all analytes with q-values $<0.05$ were 
considered significant. The data consists of significant analytes $A$, analytes in a cluster $B$, insignificant analytes $\not{A}$ and analytes not in 
the cluster $\not{B}$. The contingency table ($T_{ij}$) was populated by the amount of significant analytes in the cluster ($T_{00} = \dim{(A\cap B)}$), 
significant not in the 
cluster ($T_{01}=\dim{(A\cap\not{B})}$), insignificant in the cluster ($T_{10}=\dim{(\not{A}\cap B)}$) and all non significant analytes not in the cluster 
($T_{11} = \dim{(\not{A}\cap\not{B})}$). The top cluster transcripts were further analysed using the \ac{string-db}~\cite{szklarczyk_string_2021} for context.

We did not train a model to infer digits depending on the standardised image data but only describe the cluster content of the optimal solution. 
The MNIST data clusters were subjected to compositional analysis and benchmarked with the $\gamma$~metric 
(Equation~\ref{eq:specificity}).

\section{Results} \label{sec:results}

We confirmed that \ac{CCA} were numerically equivalent to \ac{LCA} employing single linkage construction. The 
\ac{DBSCAN} method without point rejections produced identical clustering solutions with equivalent clustering labels for all distances checked 
as compared to suggested with (Algorithm~\ref{alg:connectivity}) and \ac{LCA} (with single linkage) for the water system.


We summarize the optimisation results from the GRS~(Algorithm~\ref{alg:grs}) employing \ac{CCA} in Table~\ref{tab:optims}.

\begin{table}[t]
\begin{tabular}{lllllll}
			& $G_{max}$	& $c_{max}$	& $G_{mean}$	& $c_{mean}$	& $G_{min}$	& $c_{min}$	\\
Water 			& 1.782		& 17		& 1.836		& 11		& 1.132		& 33       	\\
Pima  			& 0.0630	& 1313		& 0.0961	& 110		& 0.1688	& 3		\\ 
MNIST			& 0.04713	& 3388		& 0.09573	& 188		& 0.45622	& 6		\\ 
\end{tabular}
\caption{ Optimal $\epsilon$ values as determined via \ac{CCA}~\ac{GRS}~Algorithm~\ref{alg:grs} }
\label{tab:optims}
\end{table}

\subsection{The Water coordinates}

\begin{figure*}[t!]
  \centering
  \begin{subfigure}[t]{0.45\textwidth}
    \centering
    \includegraphics[scale=0.35]{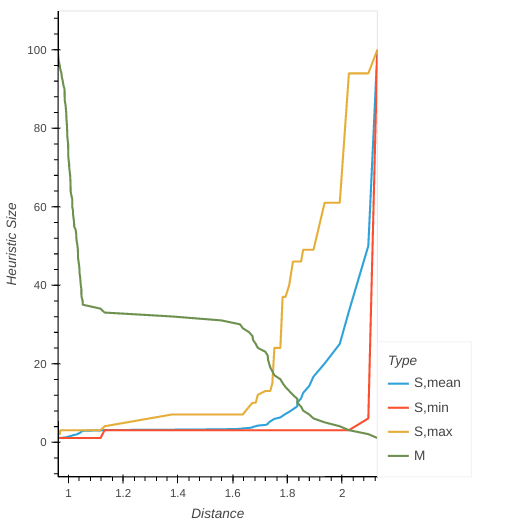}
    \caption{
	The heuristics described by Eq.~\ref{eq:M} and Eq.~\ref{eq:S} where the $S,mean$ corresponds to $\overline{S}$ while 
	$S,min$ and $S,max$ corrsponds to $S^-$ and $S^+$ respectively for different $\epsilon$ distance values. 
    }
    \label{fig:waterSM}
  \end{subfigure}
  ~
  \begin{subfigure}[t]{0.45\textwidth}
    \centering
    \includegraphics[scale=0.35]{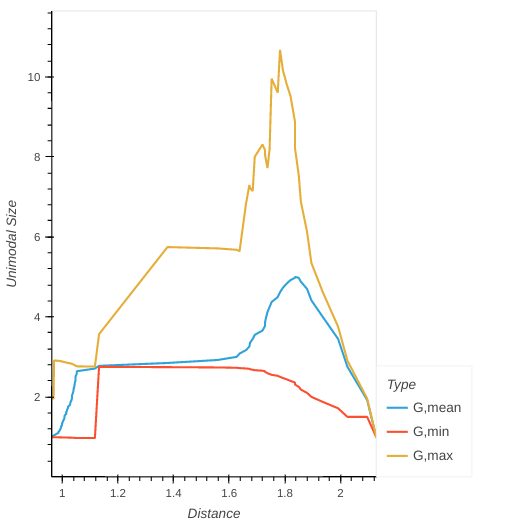}
    \caption{
	The $G$ function values for different $\epsilon$ distances. The geometric $G,label$ values have been calculated using their corresponding $S,label$
	value heuristic.
    }
    \label{fig:waterG}
  \end{subfigure}
  \caption{ 
	Overview of all distances in the \ac{AHC} single linkage solution and their corresponding value functions employed in the 
	clustering optimisation of 100 atoms belonging to water and hydronium molecules.
  }
  \label{fig:water}
\end{figure*}

In Figure~\ref{fig:waterSM} we note that the $S_{min}$ heuristic increases early and persists through much of the interval. This causes the $G_{min}$ 
function in Figure~\ref{fig:waterG} to obtain an early max shifted closer to the smallest distances than the mean. Liquid molecular water is structured 
and forms hydrogen bonds with its four tetrahedral 
neighbouring water molecules at distances smaller than $2$~[\AA]. The atomic hydrogen is bound to the oxygen at distances smaller than $1.1$~[\AA] and 
we expect hydrogen reactions between different water molecules to occur in the range $\epsilon\in[1,2]$~[\AA]. We
observe that our results, in Table~\ref{tab:optims}, for the optimal clustering solutions are all in this range. It is also clear from the $M$ heuristic in, 
Figure~\ref{fig:waterSM}, that going to larger distances would cause the entire system to become connected in a single cluster. 

Using the $S_{min}$ metric our unimodal function retains a single water molecule cluster as the heuristic reference through most of the $\epsilon$
search range and is the reason for the early jump in the $S_{min}$. The $M$ heuristic is also flat in a large range starting from the mean O-H 
position until the first atom species position of the first coordinating water molecule. This causes the $G_{min}$ to obtain its 
extremum early and clustering at this distance $\epsilon\approx 1.13$ causes the system to decompose into 
$32$ complete water molecules and a single Hydronium molecule. 

Both the $S_{max}$ and the $S_{mean}$ solutions obtain similar $\epsilon\approx 1.8$ and form cluster sizes of $17$ and $11$ segments respectively. 
One corresponds to the Hydronium centred cluster with 3 and 6 coordinating water molecules respectively as well as several smaller clusters with 
fewer water molecules. This is the expected outcome since the polarising Hydronium ion will cause a locally denser 
liquid medium in its first coordination shell~\cite{tjornhammar_molecular_2010}.

\subsection{The Pima data}

\begin{figure*}[t!]
  \centering
  \begin{subfigure}[t]{0.45\textwidth}
    \centering
    \includegraphics[scale=0.35]{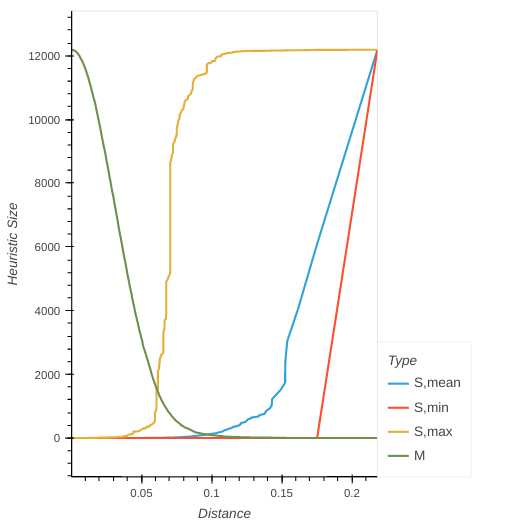}
    \caption{
        The heuristics described by Eq.~\ref{eq:M} and Eq.~\ref{eq:S} where the $S,mean$ corresponds to $\overline{S}$ while
        $S,min$ and $S,max$ corrsponds to $S^-$ and $S^+$ respectively for different $\epsilon$ distance values.
    }
    \label{fig:pimaSM}
  \end{subfigure}
  ~
  \begin{subfigure}[t]{0.45\textwidth}
    \centering
    \includegraphics[scale=0.35]{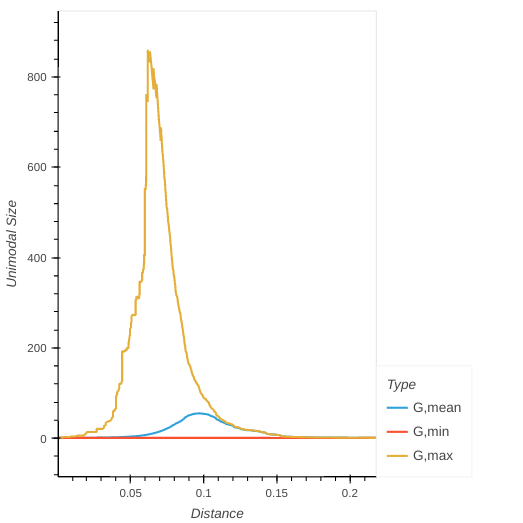}
    \caption{
        The $G$ function values for different $\epsilon$ distances. The geometric $G,label$ values have been calculated using their corresponding $S,label$
        value heuristic.
    }
    \label{fig:pimaG}
  \end{subfigure}
  \caption{
        Overview of all distances in the \ac{AHC} single linkage solution and their corresponding value functions employed in the 
        clustering optimisation of 12185 gene transcripts belonging to lean and obese men and women.
  }
  \label{fig:pima}
\end{figure*}

The optimisation protocol was evaluated for $S^+,\overline{S}$~and~$S^-$ generating successively fewer $c$, see Figure~\ref{fig:pimaG}. 
The $\min$ optimisation generated a single cluster 
with $12183$ transcripts and two smaller clusters with a single transcript in each. This rendered the decomposition uninformative. The optimisation result was 
caused by the outlier transcripts causing the persistent single component 
clusters to form. The fast decay of the total number of clusters causes the $G_{min}$ to obtain a late minimum extremum. The other two solutions, $S^+$ 
and $\overline{S}$ both obtain early optima with a larger number of clusters. The $\overline{S}$ solution contains a single huge cluster and several smaller
single or few transcript clusters. The $S^+$ solution contains several larger clusters and many trailing few component clusters. The optimal clustering 
results are visible in Figure~\ref{fig:umapped}. 

Since the microarray data has annotated groups, for all the samples, we assessed how well the cluster formation corresponded to traditional \ac{ANOVA} results.
While the clustering metrics by themselves clearly show that the useful solutions correspond to the $S^+$ and $\overline{S}$ solution. The evaluation of
enrichment for significant transcripts explaining the lean-obesity variation showed that the $S^+$ had the largest amount of significant clusters. The 
$\overline{S}$ dependent solution results in $1$ significant (q-value~$ <0.05$) cluster with a huge amount of transcripts while the $S^+$ solution 
correspond to $15$ significant clusters with sizes in the range $[10,400]$. The cluster with the highest significance contains $296$ transcripts. The
corresponding protein coding activity can be assessed via \ac{string-db} and relate to: extracellular response to stimulus and organic substances, 
rheumatoid arthritis~\cite{ashburner_gene_2000} as well as abnormal MAPK/ERK pathway signalling~\cite{kanehisa_kegg_2021}, which can lead to uncontrolled 
growth~\cite{downward_targeting_2003}. Changes to cytokine signalling, inflammation and immune 
system response~\cite{gillespie_reactome_2022} were also enriched for the cluster. These results are in line with common knowledge of the expressions of 
obesity~\cite{monteiro_chronic_2010} and prior knowledge for this data set~\cite{lee_microarray_2005}.

\subsection{The MNIST digits}

\begin{figure*}[t!]
  \centering
  \begin{subfigure}[t]{0.45\textwidth}
    \centering
    \includegraphics[scale=0.35]{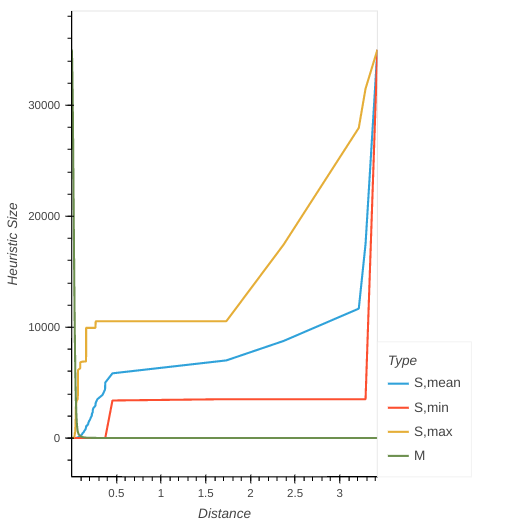}
    \caption{
        The heuristics described by Eq.~\ref{eq:M} and Eq.~\ref{eq:S} where the $S,mean$ corresponds to $\overline{S}$ while 
        $S,min$ and $S,max$ corrsponds to $S^-$ and $S^+$ respectively for different $\epsilon$ distance values.
    }
    \label{fig:mnistSM}
  \end{subfigure}
  ~
  \begin{subfigure}[t]{0.45\textwidth}
    \centering
    \includegraphics[scale=0.35]{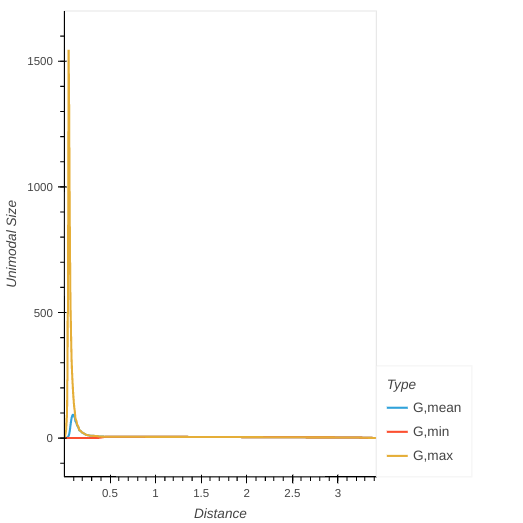}
    \caption{
        The $G$ function values for different $\epsilon$ distances. The geometric $G,label$ values have been calculated using their corresponding $S,label$
        value function.
    }
    \label{fig:mnistG}
  \end{subfigure}
  \caption{
        Overview of all distances in the \ac{AHC} single linkage solution and their corresponding value functions employed in the
        clustering optimisation of $35000$~grayscale~$28 \times 28$~pixel images from the \ac{MNIST} digits data.
  }
  \label{fig:mnist}
\end{figure*}

\begin{figure*}[t!]
  \centering
  \begin{subfigure}[t]{0.45\textwidth}
    \centering
    \includegraphics[scale=0.35]{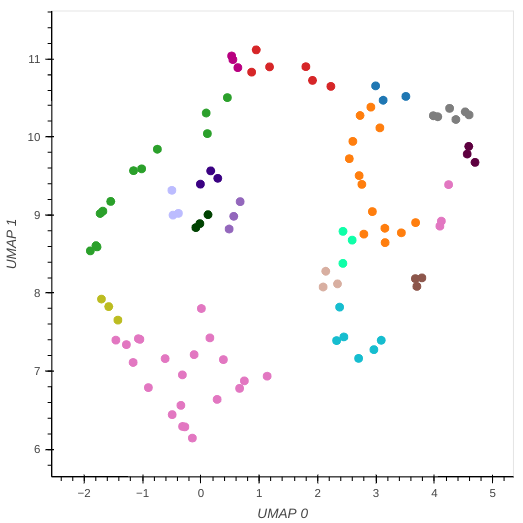}
    \caption{
        Water optimum solution employing $G_{max}$ calculated with $S^+$
    }
    \label{fig:umapWater}
  \end{subfigure}
  ~
  \begin{subfigure}[t]{0.45\textwidth}
    \centering
    \includegraphics[scale=0.35]{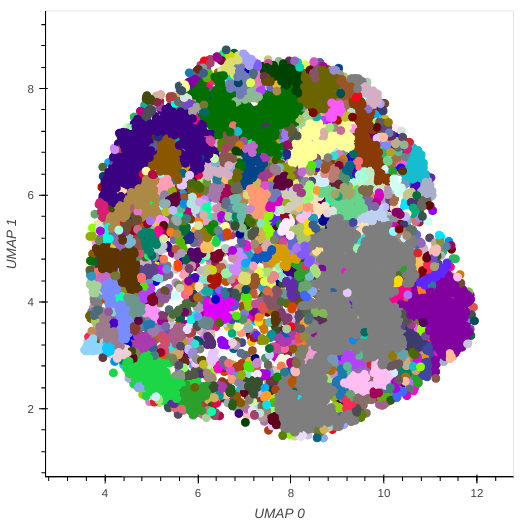}
    \caption{
        Pima optimum solution employing $G_{max}$ calculated with $S^+$
    }
    \label{fig:umapPima}
  \end{subfigure}
  ~
  \begin{subfigure}[t]{0.45\textwidth}
    \centering
    \includegraphics[scale=0.35]{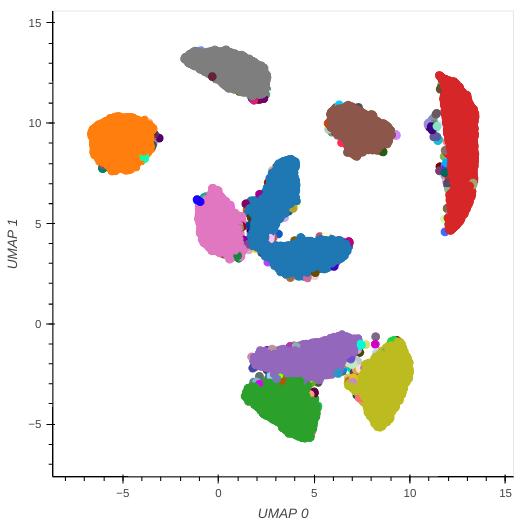}
    \caption{
        MNIST optimum solution employing $G_{mean}$ calculated with $\overline{S}$
    }
    \label{fig:umapMNIST}
  \end{subfigure}
  \caption{
        Overview of the clustering solutions for the $G$ optimum for the three data sets.
  }
  \label{fig:umapped}
\end{figure*}

The \ac{UMAP} transformation of the standardised \ac{MNIST} images obtains a clear structure, see Figure~\ref{fig:umapMNIST}, belonging to the $G,mean$ 
solution in Figure~\ref{fig:mnistG}. 

The $S^+$ dependent solution forms a large number of clusters with high specificity to specific targets, as calculated with Equation~\ref{eq:specificity}. 
The top $20$ most specific clusters, exhibiting specificity above $92\%$, all contain around $100$ to $1000$ images each but are also redundant in that 
several digits 
reappear in several clusters and that the digit $4$ is missing. It appears at $\gamma=0.86$ in a cluster with 237 images. The downside of this solution 
is a large number of clusters with $\gamma < 0.5$ which corresponds to $30\%$ of the entire data set.

Both of the solutions corresponding to the choices $S^-$ and $\overline{S}$ are more similar as compared to the $S^+$ dependent solution. 
The $S^-$ dependent solution forms $6$ clusters, see Table~\ref{tab:optims}. Of which $4$ has a specificity of over $96\%$ corresponding to the 
digits $0,1,2,6$. The remaining $2$
clusters have specificities of $34\%$ and $35\%$ and are comprised of the digits $4,7,9$ and $3,5,8$ respectively. The low specificity of the 
two larger mixed clusters means that almost $60\%$ of the data segmentation can be assumed to be unreliable.

For the $\overline{S}$ solution the number of clusters is large but there are only $13$ clusters with specificity above $50\%$ of which $9$ have sizes
above $3000$ images each. The largest cluster has obtained the lowest specificity of $51\%$ and is dominated by the digits $5,8$. The remaining clusters
contain the digits $0,1,2,3,4,6,7,9$ with specificities above $93\%$. The data set clustering with $\overline{S}$ is more exhaustive in that less than 
$1\%$ of the data has specificity lower than $50\%$.

\section{ Conclusion and Result summary }

In this work, we evaluated different metrics for unimodal optimisation for determining cluster segments. All the studied metrics resulted in 
connectivity-based clustering results that corresponded to different distance cuts through a \ac{AHC} using single linkage. The optimisation strategy 
determined the global extremum in all the tested cases. The optimisation method suggests a distance cutoff using only the metrics from the
distance matrix and the connectivity clustering results.

The choice for the $S$ function value has got a large influence on the optimal solution. For the water system using $S^-$ resulted in segments 
corresponding to the molecules in the system while the larger $\overline{S}$ and $S^+$ resulted in chemical environments consisting of several complete 
molecules. 

The remaining data sets both had labels corresponding to either feature targets or sample descriptor labels for the MNIST and Pima data sets respectively. 
This facilitated further evaluation of the usefulness of the clustering solutions. The MNIST data obtained the most informative clustering solution using the
$\overline{S}$ heuristic. The Pima data set is the densest and obtained significant and informative clusters by employing the $S^+$ heuristic.

\section{Discussion} \label{sec:discuss}

For denser data, the $S$ heuristic function probably needs to correspond to larger quantile value functions to yield a useful decomposition. For 
microarray and \ac{RNA} sequencing data the $S^+$ or a high quantile value function is the suggested heuristic for finding a useful optimal $\epsilon$ cut.
This method will not work well with highly structured data, where the system is entirely connected at one distance, but functions as expected when the data
is connected via an ensemble of distances. Then finding a useful cluster representation for visualizing the data becomes a simple optimisation task.

\newpage
\bibliography{comc.bib}

\end{document}